\documentclass{osa-article}

\journal{oe}

\begin{document}

\title{Imaging through highly scattering environments using ballistic and quasi-ballistic light in a common-path Sagnac interferometer}

\author{Jesse Dykes,\authormark{1} Zeina Nazer,\authormark{1} Allard P. Mosk,\authormark{2} and Otto L. Muskens\authormark{1,*}}

\address{\authormark{1}Physics and Astronomy, Faculty of Engineering and Physical Sciences, University of Southampton, Southampton, UK\\
\authormark{2}Utrecht University, Debye Institute for Nanomaterials Science, The Netherlands\\}

\email{\authormark{*}O.Muskens@soton.ac.uk} 



\setcopyright{}

\begin{abstract}
The survival of time-reversal symmetry in the presence of strong multiple scattering lies at the heart of some of the most robust interference effects of light in complex media. Here, the use of time-reversed light paths for imaging in highly scattering environments is investigated. A common-path Sagnac interferometer is constructed which is able to detect objects behind a layer of strongly scattering material through up to 14 mean free paths total attenuation length. A spatial offset between the two light paths is used to suppress non-specific scattering contributions, limiting the signal to the volume of overlap. Scaling of the specific signal intensity indicates a transition from ballistic to quasi-ballistic contributions as the scattering thickness is increased. The characteristic frequency dependence for the coherent modulation signal provides a path length dependent signature, while the spatial overlap requirement allows for short-range 3D imaging. The technique of common-path, bistatic interferometry offers a conceptually novel approach which could open new applications in diverse areas such as medical imaging, machine vision, sensors, and lidar.
\end{abstract}

\section{Introduction}
The development of new methods capable of detecting objects on length scales from millimeters to hundreds of meters is of great importance for technology areas ranging from biomedical imaging to autonomous vehicle navigation. Commonly used methods include optical coherence tomography, time-of-flight and coherent lidar techniques \cite{Berkovic2012, Dunsby2003}. Most imaging and ranging methods are strongly challenged in performance in situations of low visibility due to haze and scattering when dealing with environments such as biological tissue, rain, snow, fog, and clouds \cite{Dunsby2003, Tobin2019, Feneyrou2017}. In such environments, the signal of interest is strongly attenuated while non-specific scattering backgrounds are greatly enhanced. Techniques that are robust against strong scattering are of interest to increase the visibility and hence improve safety, e.g. in autonomous vehicles.

In scattering environments, ballistic and quasi-ballistic components are often the only recoverable specific signatures, and efforts at extending the detection range of these components are of importance. In optical coherence tomography (OCT), recovery of ballistic signals is possible through many mean free paths while the imaging distances are generally on the millimeter scale \cite{Fer2003, Nguyen2013}. On larger length scales, time-gated photon counting is another well-explored strategy, of particular interest for pulsed lidar \cite{Yoo1990,Wan1991,Tobin2019}. Next to the recovery of weak ballistic components, detection and imaging of objects using the scattered light itself is seeing seen rapidly increasing interest \cite{Yaqoob2008, Pop2010, Bertol2012, Kat2012, Kanaev2018,Badon2016}. While significant progress is being made to speed up reconstruction of scattered information \cite{Tzang2019}, in many dynamic scattering environments the decorrelation times are still too short for speckle-based reconstruction methods.

A specific class of methods for discriminating scattered and ballistic components involves bistatic techniques where illumination and detection paths are spatially separated to allow rejection of scattering. Spatially offset detection has been used particularly successfully in Raman detection through opaque media \cite{Matousek2005, Chen2018}. In OCT, the use of spatially offset, bistatic detection has been shown to provide an additional advantage for the selection of signals from deep inside scattering tissue \cite{Matthews2014, Zhao2017}. In long-range bistatic lidar, similar principles of spatially offset excitation and detection paths are used to distinguish signals from different atmospheric regions \cite{Barnes2007}. These techniques have in common that only a single, unidirectional path from illumination to detection is used.

In this work, we propose and demonstrate a conceptually distinct approach to active detection of objects in scattering environments. Our approach is motivated by the guiding principle that time-reversed light paths produce some of the most robust interference effects in scattering media \cite{Albada1985,Wolf1985} and therefore these paths could be useful for detection and imaging of objects. An example of this idea is low-coherence enhanced backscattering spectroscopy, where interference of time-reversed light paths in backscattering is used for depth resolved biomedical imaging \cite{Boustany2010, Kim2006}. Time-reversal symmetry is also a prominent feature of the common-path Sagnac interferometer \cite{Sagnac1913}, which has found wide application in fiber optic gyroscopes \cite{Vali1976, Chow1985}. We furthermore highlight the exceptional phase stability of common-path interferometers, with applications for example in metrology \cite{Hurley1999, Sugawara2002} and gravitational wave detection \cite{Sun1996}.

In our studies we explore the use of common-path interferometry for detection of a target after attenuation by a layer of multiple scattering material. To successfully retrieve an exponentially reduced specific signal from a large nonspecific background, it is necessary that the specific signal can be distinguished by a unique signature. The interference of time-reversed light paths in combination with an electro-optic modulator to encode a phase modulation provides such a particular signal and allows sensitive detection after many attenuation lengths through the scattering medium. It is found that the scaling of target signal with attenuation length follows a less than exponential trend, indicating that the technique, next to providing the ballistic component, has a sensitivity to non-ballistic scattered light travelling in the same direction as the ballistic beam. The use of this signal for imaging and ranging is critically discussed.

\section{Method}

\begin{figure}[t!b]
	\centering
	\includegraphics[width=12cm]{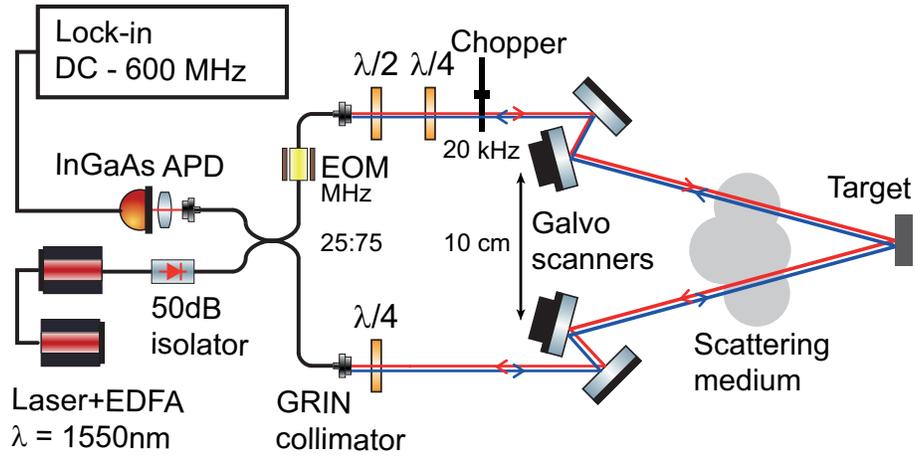}
	\caption{ Optical setup, including fiber optical system with laser at 1550~nm wavelength, erbium-doped fiber amplifier (EDFA), optical isolator, 75:25 splitter, electro-optic modulator (EOM), and graded-index (GRIN) collimators, and free-space optics with waveplates ($\lambda/2$, $\lambda/4$), optical chopper, galvo scanners and InGaAs avalanche photodetector (APD). Blue and red lines indicate the two counterpropagating light paths to the target and through the scattering medium (grey area).}
	\label{Fig: Opticsetup}
\end{figure}

The experimental arrangement is shown in Figure~\ref{Fig: Opticsetup}. The method makes use of a common-path Sagnac interferometer consisting of two counterpropagating light paths, where the target object forms part of the interferometer. A 1550~nm narrowband laser (Keysight) with a bandwidth of 500~kHz was amplified using a polarization maintaining Erbium-doped Fiber Amplifier (Pritel) to a total power of 400~mW. An optical isolator was used to protect the laser against backscattering from the setup. The output was split into two arms using a 75\%:25\% fiber optic splitter, of which the 25\% port was sent to an electro-optic phase modulator (EOM, Jenoptik) and 75\% was coupled out directly to free space. This unbalanced ratio of the splitter was chosen to reduce the power through the EOM to avoid damage to the component. Polarization-maintaining (PM) components were used throughout the fiber optic system, with angled (APC) fiber couplers to minimize spurious reflections. The two outputs were collimated using graded-index (GRIN) lensed fibers with specified return loss less than -60~dB (Oz Optics). Despite the use of fiber components with very low return loss, internal reflections still provided a significant background far above the detector noise floor. Therefore, an additional low-frequency (20~kHz) chopper in the free space path was included to reject spurious signals originating from the fiber optical system. A half-wavelength ($\lambda/2$) waveplate was used to align the linear polarization states of the two outputs, which is a prerequisite for obtaining an interferometric signal. Additionally, a quarter waveplate was inserted in each of the arms to reduce backscattering into the same arm, exploiting the helicity nonconserving characteristics of single backscattering events. The use of circular polarized light reduced the noise floor by a factor three, mainly due to reduced backscattering from the mechanical chopper.

The two collimated beams with 0.9~mm $1/e^2$ half width were directed to the target using a synchronized pair of 2D galvo scanners which were used for imaging at rates of 20,000 pixels per second. After interaction with the target, backscattered light was collected using the same GRIN lenses and was sent through the fiber optical system in opposite direction where it was detected at the return port of the fiber splitter using an avalanche photodiode (APD, Thorlabs). The APD had a bandwidth of 400~MHz and noise equivalent power (NEP) of 0.3 pW/Hz$^{1/2}$. The frequency modulated components were extracted from the APD signal using a high-frequency lock-in amplifier (Zurich Instruments UHFLI) with an operating bandwidth of DC-600~MHz. The phase modulation resulted in a typical Fourier spectrum comprising of harmonics of the EOM driving frequency $f$. The highest modulation amplitude was obtained at the second harmonic of the driving frequency $2f$, rather than at the fundamental $f$. This property is useful as it allows us to avoid the residual amplitude modulation of the EOM occurring primarily at $f$.

\section{Frequency dependence of common-path signal}

\begin{figure}[t!b]
	\centering
	\includegraphics[width=10cm]{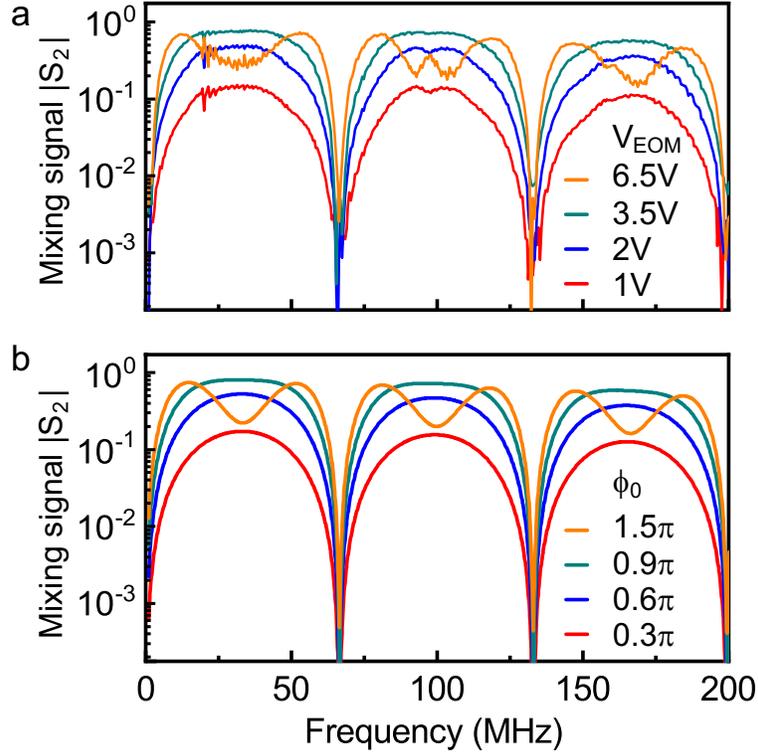}
	\caption{Measured (a) and calculated (b) modulus of the mixing signal $|S_2|$ at 2$f$ against driving frequency $f$, for a fixed roundtrip time $\tau=5$~ns, corresponding to 1.5~m path length. Different values of the EOM driving voltage correspond to different peak-to-peak phase amplitudes $\phi_0$, shown by calculated curves in (b).}
	\label{Fig: fsweep}
\end{figure}

The mixing of scattered fields collected from the two counterpropagating light paths provided a phase stable interferometric signal under the condition that the system is stable during the roundtrip time and is not rotating. The EOM breaks the symmetry by introducing a time-varying phase $\phi(t)=\phi_0/2 \sin (2 \pi f t)$, where $\phi_0$ denotes the peak to peak amplitude of modulation and $f$ is the EOM driving frequency. Light travelling through the two counterpropagating light paths arrives at the EOM from opposite sides, separated by a time delay $\tau$. The mixing of these two scattered field components results in a heterodyne mixing signal $S(t)=E_{\rm CW} E_{\rm CCW}^* =|E|^2 \exp[i(\phi(t+\tau)-\phi(t))]$ where $|E|$ denotes the amplitude of the two scattered fields which is taken to be equal by symmetry. The derivation of the power spectrum from the phase-modulated signal follows standard trigonometric manipulations similar to diffraction by sinusoidal phase gratings \cite{Beach1997, Bergh1981, Malykin2007}. After expansion of terms $\exp[i \phi_0 \sin(2 \pi f t)/2 ]= \sum_{q=-\infty}^{\infty} J_q(\phi_0/2) \exp[i 2\pi q f t]$ and following a Fourier transform, the heterodyne mixing signal is obtained as
\begin{equation}
S(\omega)=\frac{|E|^2}{2\pi} \sum_{q,p=-\infty}^\infty J_{q}(\phi_0/2) J_{p}(\phi_0/2) \exp{(i 2 \pi q f \tau)} \delta[\omega-2 \pi (q-p) f] \, .
\end{equation}
Identifying $n=q-p$ as the harmonic of the frequency spectrum defined by the Dirac delta function, we obtain an expression for the $n$-th harmonic components, $S_n$, according to
\begin{equation}
S_{n}=\frac{|E|^2}{2\pi} \sum_{q=-\infty}^\infty J_{q}(\phi_0/2) J_{q-n}(\phi_0/2) \exp{(i 2 \pi q f \tau)} \, .
\end{equation}
The modulus of each harmonic is then obtained through the spectral power, $P_n=|S_n|^2$, which follows as
\begin{eqnarray}
P_{n} &=& \frac{|E|^4}{4\pi^2} \sum_{q,r=-\infty}^\infty J_{q}(\phi_0/2) J_{q-n}(\phi_0/2) \nonumber \\
&\times & J_{r}(\phi_0/2) J_{r-n}(\phi_0/2) \cos[2 \pi f \tau (q-r)] \, . \label{Eq: harmonic}
\end{eqnarray}
Figure~\ref{Fig: fsweep}a shows experimental frequency sweeps of the modulus of the second harmonic signal intensity $|S_{2}|=(P_2)^{1/2}$ for four different values of the EOM driving voltage. Fits to the data are given in Figure~\ref{Fig: fsweep}b and show good agreement with the general behaviour for values of $\phi_0$ in the range $0.3 \pi$ to $1.5 \pi$. Here, the extraction of the modulated signal using the lock-in amplifier was done using a two-stage scheme where the output of the signal at the harmonic of the EOM drive frequency $f$ was measured with a short (4~$\mu$s) integration time, and subsequently the modulus of this signal was used as the input for the second lock-in stage operating at the chopper frequency $f_{\rm chop}$. This method was suitable for the lock-in amplifier to track the EOM frequency sweep. Alternatively, direct detection at the sum frequency of the EOM and chopper was possible for a fixed modulation frequency. From the expression for $I(t)$ we immediately identify that for $\phi(t+\tau)=\phi(t)$, the phases are the same and modulated signal is zero. This condition is fulfilled for frequencies equal to an integer multiple $n$ of the  inverse of the roundtrip time, $f=n/\tau$. From the zeros in the frequency sweep, we find a roundtrip distance of 4.5~m, which includes approximately 2~m length of fiber from the EOM to the GRIN output coupler and 1~m of roundtrip distance to the target.

\section{Object detection and imaging through scattering media}

To investigate the attenuation of the target signal after propagation through a scattering medium, we measured the signal after propagation through a stack of sheets of Polytetrafluoroethylene (PTFE, Teflon), where each sheet was 0.5~mm thick. The weakly scattering PTFE is considered a reasonable approximation for light scattering in biological tissue, while it is a simplification for environmental scattering media like clouds and fog which are otherwise considerably more difficult to model in a laboratory environment. The characterization of the scattering medium is presented in the Appendix. Using ballistic and total transmission measurements we obtained a scattering mean free path at 1550~nm wavelength of $l_s=0.36 \pm 0.02$~mm, a transport mean free path $\ell=0.50 \pm 0.03$~mm, and a diffuse absorption length $L_a=10 \pm 1$~mm for PTFE. Each slab therefore represents a total attenuation length in reflection (i.e. double pass) of $2L/l_s=2.8$. Our values of the scattering length are three times larger than values reported at 800~nm wavelength for the same material \cite{Wiersma2000}, which is consistent with the expected wavelength scaling.

\begin{figure}[t!b]
	\centering
	\includegraphics[width=10cm]{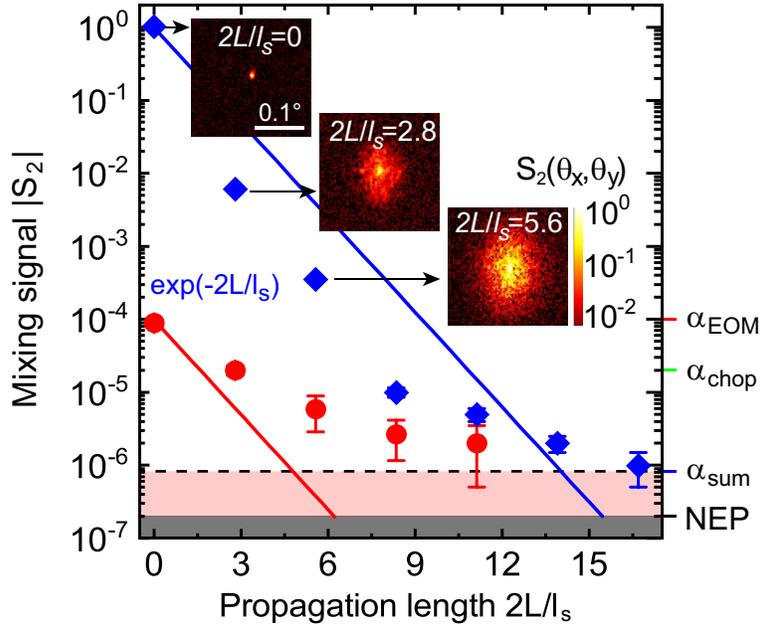}
	\caption{Dependence of reflection signal intensity $|S_2|$ on total attenuation length $2L/l_s$, for the specular reflection from a protected silver mirror (dots, blue) and diffuse reflector (diamonds, red, DG10-220, Thorlabs). Blue and red lines indicate exponential decay using Beer-Lambert law $|S_2| \propto \exp{(-2 L/l_s)}$. Lines on right indicated background (noise) levels for EOM only at $2f$ ($\alpha_{\rm EOM}$) and chopper only at $f_{\rm chop}$ ($\alpha_{\rm chop}$ modulation, as well as sum frequency detection at $2f+f_{\rm chop}$ ($\alpha_{\rm sum}$), with NEP indicating the detector noise floor.  Images in inset: angular $\theta_x, \theta_y$ scans (log-scale) of the target showing angular spread of detection signal for attenuation lengths of 0, 2.8, and 5.6 mean free paths.}
	\label{Fig: slabs}
\end{figure}

The target signal intensity was measured for increasing thicknesses of the scattering PTFE stack of up to 6 slabs ($2L/l_s=16.8$). Figure \ref{Fig: slabs} shows the mixing signal intensity against scattering length using a mirror as a target with high specular reflectivity (diamonds, blue). The mirror target was placed at a distance of 50~cm from the galvo scanners. For the cases of zero and one PTFE slab, additional neutral density filters were used to attenuate the signal. All values are normalized to the return signal obtained without scattering medium. The figure also shows detection limits when using only the EOM modulation ($\alpha_{\rm EOM}$), which is limited by internal reflections in the fiber, or for only the mechanical chopper ($\alpha_{\rm chop}$), which is limited by low-frequency (kHz) noise. The combined modulation scheme at the sum of the EOM and chopper frequencies ($\alpha_{\rm sum}$) is substantially lower and is found to be ultimately limited by backscattering from the chopper in our setup, while the sensitivity limit of the APD (NEP) is still a factor of five lower than this limit.

In our test setup, a discernible signal could be detected at up to 14 mean free paths of total attenuation length. This detection limit corresponds to a dynamic range of 6 orders of magnitude in intensity. An additional order of magnitude dynamic range could be gained by placing $\times 3$ beam expanders in the arms, however this only affected the throughput of the collimated beam in absence of scattering (i.e. the highest detected intensity) which was limited by beam divergence over the free-space roundtrip, whereas the lowest signals were limited by the total \'etendue (angle times area) of the system which is unaltered by the beam expansion.

\begin{figure}[t!b]
	\centering
	\includegraphics[width=11cm]{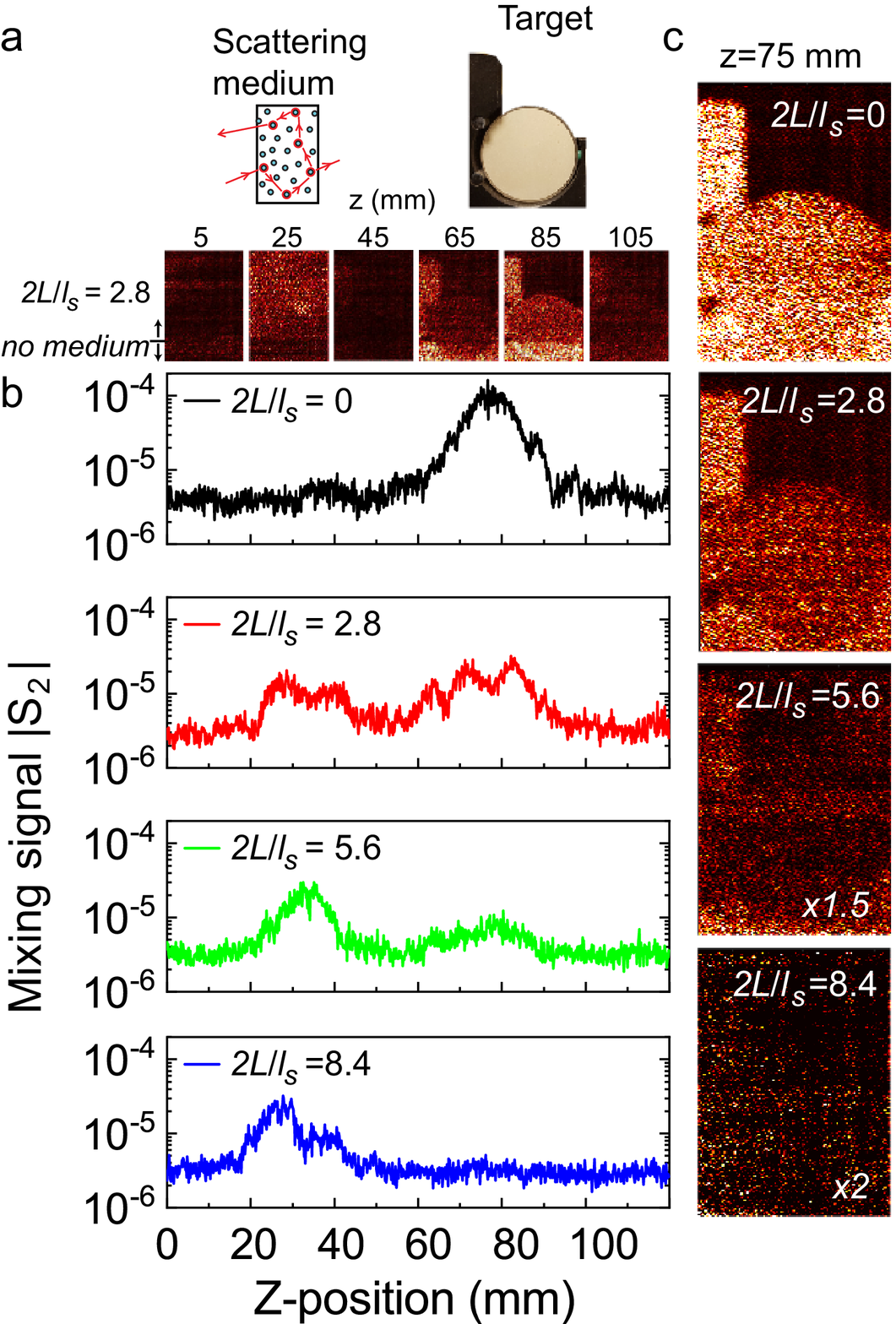}
	\caption{(a) Images taken at different positions $z$, showing position of scattering medium and target diffuse reflector. Top part of image was covered by scattering medium with $2L/l_s=2.8$, bottom part without scattering medium. (b) Dependence of mixing signal intensity $|S_2|$ (log-scale) versus position $z$ for different attenuation lengths from 0 to 8.4 mean free paths. (c) Image of target at $z=75$~mm for attenuation lengths corresponding to (b). }
	\label{Fig: zscan}
\end{figure}

It was found that the combination of surface roughness (and possibly some near-forward bulk scattering) of the stack of PTFE slabs distorted the beam profile of the ballistic beam, resulting in an angular spreading of the beam from the original diffraction-limited beam with $\theta \simeq 0.005^\circ$ to around $0.1^\circ$, as can be seen in the angular scans of the signal $|S_2|(\theta_x, \theta_y)$ for the first three points shown as insets in Figure~\ref{Fig: slabs} (maps are normalized and on logarithmic scale). The distortion results in a strongly reduced peak intensity at zero angle, below the expected attenuation $|S_2|=\exp(-2L/l_s)$ as indicated by the blue line. However, for attenuation lengths above 8 mean free paths, we see a different slope and eventually the predicted exponential attenuation even crosses below the measured signal level at $2L/l_s = 14$. The different slope indicates that the setup is able to recover some of the non-ballistic light that is produced by the surface roughness and bulk scattering and which travels in the same direction as the ballistic beam.

The non-ballistic origin of the signal is also seen when replacing the specular mirror target by a diffuse reflector (Thorlabs DG10-220), resulting in the signal indicated by the red dots in Figure~\ref{Fig: slabs}. Without the scattering medium, the diffuse target already resulted in four orders less signal intensity than the specular target. The attenuation of intensity with increasing scattering length is much less steep than the expected exponential (Lambert-Beer) law for ballistic components, suggesting that double small-angle scattering events occur that add to the coherently detected signal. We emphasize that the detection of the non-ballistic component is nontrivial as it involves a phase stable heterodyne mixing signal, which is provided by the time-reversal symmetry of light travelling in both directions through the quasi-ballistic light paths. This property renders the common-path technique more robust against scattering than other coherent detection schemes.

The presence of a specific signal from the target in the common-path interferometer can be used for imaging through the scattering medium. A gap between the target and the scattering medium 50~mm was set to provide a spatial separation of the target signal and scattering background as illustrated in Figure~\ref{Fig: zscan}a. Images were taken at different depths $z$, which could be done either by translating the scene (target plus scattering medium) or by adjusting the galvo mirrors to change the overlap in $z$, both resulting in the same response. The movie frames shows images collected at different depths $z$ for a single slab of teflon ($2L/l_s=2.8$). The bottom part of the image was not covered by the slab, hence showing the signal without scattering medium. The target is located at $z=80$~mm and the images reveal a depth of view of $\pm 10$~mm around this position. Figure~\ref{Fig: zscan}b shows the collected peak intensity in the centre of the image while the $z$-position was scanned. For increasing amounts of attenuation $2L/l_s$, the target signal intensity decreases as seen in Figure~\ref{Fig: slabs}b. The addition of the scattering medium results in a second signal located at around 30~mm, corresponding to the position of the slabs. For attenuation over 5.6 scattering lengths, the scattering of the medium overwhelms the target response, however the latter can still be distinguished because of the spatial offset of the two signals. The spatial offset of the two light paths results in an effective rejection of scattering from the individual arms, and signals are only observed from the volume in which the two beams are overlapping.

\section{Potential for 3D imaging and ranging}

\begin{figure}[tbh]
	\centering
	\includegraphics[width=9cm]{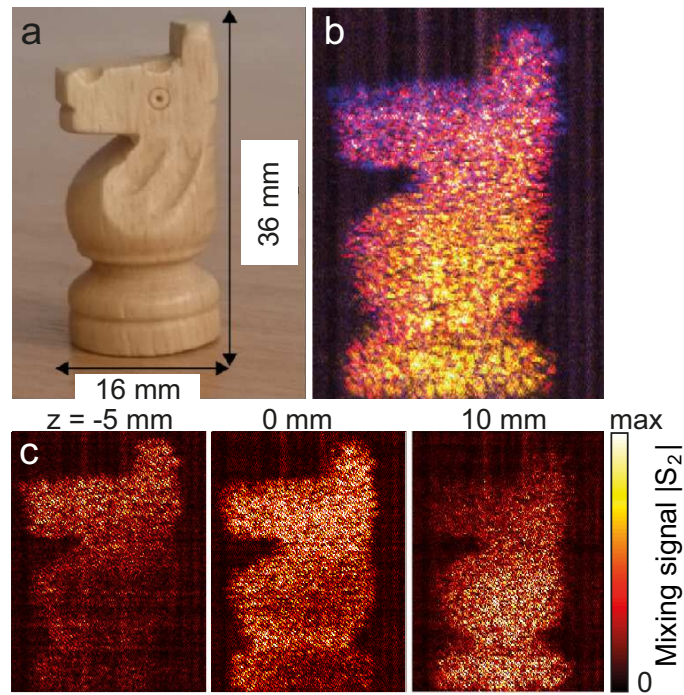}
	\caption{(a) Photograph of chess piece used in experiment. (b) Composite image of 3D chess piece using color channels blue, red, yellow for three distances. (c) corresponding images taken at relative positions $z=-5$~mm, 0~mm and 10~mm.}
	\label{Fig: 3dimage}
\end{figure}

\begin{figure}[tbh]
	\centering
	\includegraphics[width=10cm]{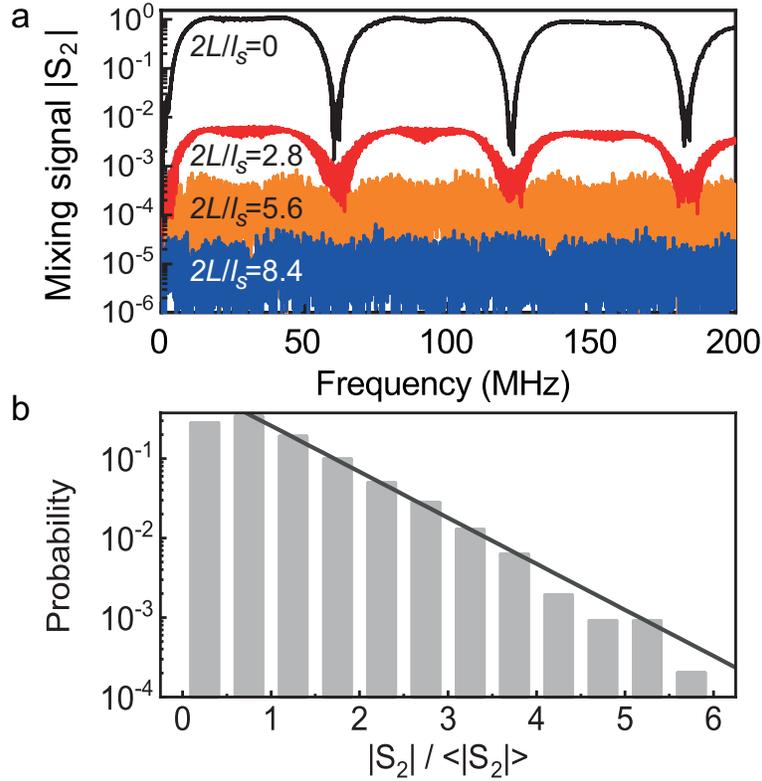}
	\caption{(a) Frequency sweeps of signal from a mirror reflector, for different attenuation lengths, showing loss of characteristic modulation signature in presence of strong multiple scattering. (b) Probability histogram of 10,000 measurements taken for the $2L/l_s=8.4$ attenuation lengths and normalized to the ensemble average $\left< |S_2| \right>$, with black line showing Rayleigh statistics for a speckle contrast of 75\%.}
	\label{Fig:sweepscatter}
\end{figure}

To demonstrate the capability of extracting three-dimensional object information, we performed measurements using a test object shown in Figure~\ref{Fig: 3dimage}a. The object, a wooden chess piece, was imaged using the common-path interferometer at three different positions $z$. A composite image is shown in Figure~\ref{Fig: 3dimage}(b) was generated by encoding the different images into three colors blue, red and yellow, while the individual images are shown in Figure~\ref{Fig: 3dimage}(c). Images taken at a separation of 5~mm in depth show different parts of the structure and reveal the three dimensional nature of the object. The depth of view is determined by the angle between the two incident beams, which amounted to 10$^\circ$ in our study. Steeper incidence can be obtained by increasing the separation of the two galvo mirror scanners, or by reducing the object distance. The composite image clearly shows a grainy structure representing individual angular speckles of the light scattered from the object back into the aperture of the fiber optic system.

The direct relation between the roundtrip time and the zeros in the response against driving frequency could be of interest for applications in ranging. This relation is explored in Eq.~\ref{Eq: harmonic} and Fig.~\ref{Fig: fsweep} for a phase stable signal from a perfect mirror and it is important to assess the robustness of this signature in the presence of scattering. Figure~\ref{Fig:sweepscatter}a shows the frequency sweeps taken for the mirror target in presence of scattering medium of up to $2L/l_s=8.4$ total attenuation length. While the signature is retained for weak scattering up to 2.8 attenuation lengths, the characteristic frequency dependence is absent for strongly scattering environments. We found in Fig.~\ref{Fig: slabs} that in this regime the ballistic beam is already distorted by surface roughness and near-forward scattering, and the angular profiles showed a grainy speckle structure. The frequency spectrum in Fig.~\ref{Fig:sweepscatter}a shows a rapid variation of intensities which can be interpreted as speckle. Indeed a histogram of the intensities taken over 10,000 points shown in Fig.~\ref{Fig:sweepscatter}b shows a characteristic exponential Rayleigh statistics, where the line represents the Rayleigh probability distribution for a speckle contrast of 75\%. Thus, the detected signal, while phase stable, shows a fast intensity modulation due to the dynamics of the speckle field. An identical behaviour was found for the diffuse reflector in absence of scattering medium (not shown here). We therefore speculate that the increased role of phase fluctuations in the near-forward speckle is responsible for the flattening of the frequency response and disappearance of the characteristic zeros for return signals $|S_2|$ below $10^{-4}$ of the signal of the reference mirror. Therefore long-distance ranging using this method appears possible for well defined reflection signals from specular targets, but is easily compromised in the presence of scattering or speckle produced by beam distortions or diffuse target reflections.

\section{Conclusion}
In conclusion, we have developed an imaging system making use of time-reversed light paths in a common-path configuration to detect objects. The properties of the technique include spatially-offset beams for rejection of single scattering, electro-optic modulation for phase-sensitive detection of the time-reversal interference, and a distant dependent ruler. We successfully demonstrate imaging of targets through a scattering medium with total attenuation lengths of up to 14 mean free paths, which can likely be further increased by improvements in instrumentation. We expect that elements of the technique will be of interest in a variety of applications in object detection and imaging, potentially including medical imaging, machine vision, sensors, and lidar. Our results invite future studies which could explore the windows of opportunity of this approach in these specific applications.

\section*{Appendix}
\subsection*{characterization of teflon scattering at 1550~nm}
\begin{figure}[t!b]
	\centering
	\includegraphics[width=10cm]{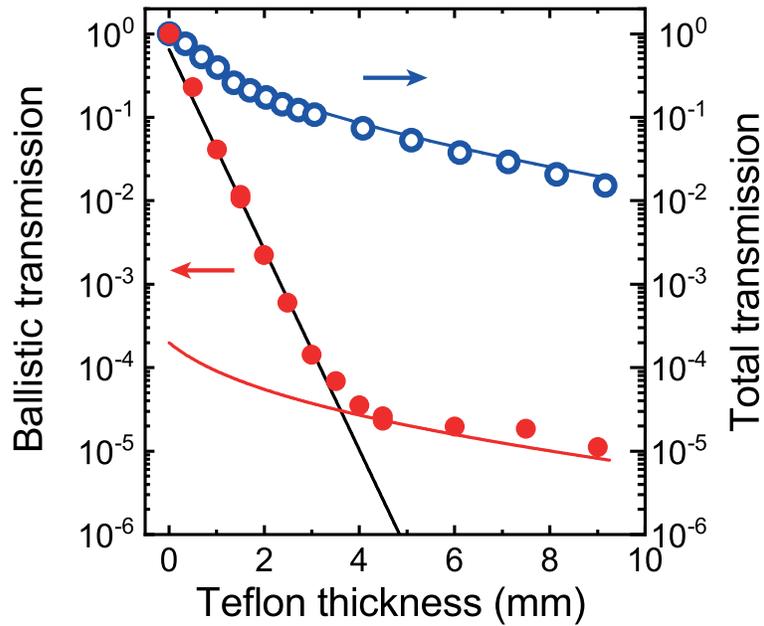}
	\caption{Ballistic transmission (red dots) and total transmission (blue open circles) through stacks of teflon slabs of increasing total thickness. Lines are fits using Beer-Lambert equation $I/I_0=\exp{(-L/l_s)}$ (black) and total transmission $T_{\rm tot}=(1+\tau_e)/(L/l_t+2 \tau_e) \exp{(-L/L_{\rm abs})}$ (blue). Red line is total transmission scaled by factor $2 \times 10^{-4}$. }
	\label{Fig: teflonstacks}
\end{figure}

The optical scattering parameters of teflon at 1550~nm wavelength were studied using a combination of ballistic beam transmission and diffuse total transmission experiments. Figure \ref{Fig: teflonstacks} shows the optical transmission through stacks of Teflon slabs, for both the ballistic component and the total transmission. The ballistic light was measured through a 1~cm diameter iris positioned at a distance of 1~m from the scattering medium, therefore collecting light transmitted within an angle of 10~mrad from the forward direction. Total transmission was measured using an integrating sphere.

For a thick medium, total transmission is determined by the diffuse scattering component given by the relation $T_{\rm tot}=(1+\tau_e)/(L/l_t+2 \tau_e) \exp{(-L/L_{\rm abs})}$. Here, $\tau_e$ is the extrapolation factor which contains internal reflection due to the index mismatch and which is close to 1 for the refractive index of teflon. Absorption is included through the diffuse absorption length which in diffusion approximation is defined as $L_{\rm abs}=(l_t l_a/3)^{1/2}$. Here $l_a$ is the bulk absorption length. A model fit using this equation (blue line in Figure~\ref{Fig: teflonstacks}) gives a transport mean free path of $0.50 \pm 0.03$~mm and a diffuse absorption length of $10 \pm 1$~mm. Thus absorption is quite low for the experimental conditions of this work.

The ballistic transmission (red dots in Figure~\ref{Fig: teflonstacks}) shows an exponential decay with the slab thickness up to about 4~mm with a scattering length of $0.36 \pm 0.01$~mm. The difference between the scattering and transport mean free paths indicates that scattering is not completely isotropic, and it takes approximately 1.5 scattering lengths to completely average over all angles. After 4~mm the slope is reduced, indicative that the diffuse scattering component dominates the transmission in the selected aperture for larger thickness. This thickness corresponds to about 10 scattering mean free paths. The red line corresponds to a fraction of the diffuse total transmission corresponding to the ratio of detected scattering solid angle ($3\times 10^{-4}$~sr) to the hemispheric total transmission.

\section*{Funding}
Defense Science \& Technology Laboratory (Dstl) (DSTLX-1000108034, DSTLX-1000127842); Engineering \& Physical Sciences Research Council (EPSRC) (EP/J016918/1).\\

\section*{Acknowledgments}
The authors would like to thank Xiaoqing Xu for his help in the initial phase of the experiment. OM acknowledges support by the University of Utrecht through the Debye Chair visiting professorship.

\section*{Disclosures}
The authors declare no conflicts of interest.

\section{References}


\end{document}